\documentclass[preprint,12pt]{elsarticle}

\usepackage{amssymb}
\usepackage[T1]{fontenc}
\usepackage[latin1]{inputenc}
\usepackage{amsmath}
\usepackage{graphicx}
\usepackage{dcolumn} 
\usepackage{bm} 
\usepackage{amsfonts}
\usepackage{epsfig}
\usepackage{dsfont}

\newcommand{\quot}[1]{``#1''}

\newcommand{\N}{{\cal N}}

\newcommand{\V}{{\cal V}}

\newcommand{\HH}{{\cal H}}

\newcommand{\Eq}[1]{Eq.~(\ref{#1})}

\begin{document}

\begin{frontmatter}



\title{Efficient classical simulation of the Gisin-Massar quantum cloning machine}


\author[label1,label3]{Hamed Saberi}
\author[label1,label2]{Yousof Mardoukhi}

\address[label1]{Department of Physics, Shahid Beheshti University, G.C., Evin, Tehran 19839, Iran}
\address[label2]{Department of Physics, Tampere University of Technology, Tampere, Finland}
\address[label3]{School of Physics, Institute for Research in Fundamental Sciences (IPM), Tehran 19395-5531, Iran}

\begin{abstract}


We provide here the technical details of the recently proposed tensor-network protocol for classical simulation of the Gisin-Massar quantum cloner by the authors [Phys. Rev. A, \textbf{85}, 052323 (2012)]. The protocol essentially instructs how to efficiently simulate an optimal quantum cloning machine of Gisin-Massar on a classical computer. A viable computational platform for generation on demand of arbitrary number of optimal clones with controllable numerical resources is realized by rephrasing the Gisin-Massar output state in terms of the hierarchy of the so-called matrix-product states (MPS) and invoking parity features peculiar to such an output.

\end{abstract}

\begin{keyword}
optimal quantum cloning \sep Gisin-Massar state \sep matrix-product state (MPS) \sep singular-value decomposition

\end{keyword}

\end{frontmatter}

\section{Introduction}
\label{sec:intro}

The linearity of quantum mechanics prohibits replication of an arbitrary and \emph{a priori} unknown quantum state~\cite{Wootters1982}. More precisely, the requirement for making a clone of an unknown arbitrary state will be penalized by an imperfect (less than unity) \quot{fidelity} as a measure of the quality of the copy. Nevertheless, approximate or probabilistic quantum cloning will be essential for various quantum informational tasks and applications. This has consequently stimulated a great deal of research for designing quantum cloning machines (QCM)~\cite{Scarani2005,Chang2013,Fan2013} that provide a recipe for producing approximate quantum clones with an \emph{optimal} fidelity in the sense that the cloned states exhibit the maximum possible similarity to the original input state. Among them, we consider here the symmetric universal quantum cloning machine (UQCM)~\cite{Buzek1996} of Gisin and Massar~\cite{Gisin1997} according to which the state-independent cloning of a single qubit in an arbitrary unknown input state $|\psi_{\mathrm{in}}\rangle \equiv \alpha|0\rangle + \beta|1\rangle$ to identical $M$ clones is described by an isometry map $\hat{\V}_{1 \rightarrow M}$ of the form
\begin{eqnarray}
\label{eq:Gisin-Massar}
|\psi_{\mathrm{in}}\rangle \rightarrow |\Psi_{\mathrm{out}}\rangle & = & \hat{\V}_{1 \rightarrow M} |\psi_{\mathrm{in}}\rangle = |GM_{M}(\psi_{\mathrm{in}})\rangle\\
\nonumber
&\equiv & \sum_{j=0}^{M-1} \gamma_{j}|(M-j)\psi_{\mathrm{in}},j \psi_{\mathrm{in}}^{\perp} \rangle_{S}
\otimes |(M-j-1)\psi_{\mathrm{in}}^{a},j \psi_{\mathrm{in}}^{a \perp} \rangle_{S}  \; ,
\end{eqnarray}
with the coefficients
\begin{eqnarray}
\label{eq:gamma_j}
\gamma_j \equiv \sqrt{{2(M-j) \over M(M+1)}}  \; ,
\end{eqnarray}
and $|(M-j) \psi_{\mathrm{in}}, j \psi_{\mathrm{in}}^{\bot}\rangle_S$ denotes the normalized completely symmetric state (under all possible qubit permutations, e.g., $|\psi_1, \psi_2\rangle \to |\psi_2, \psi_1\rangle$ for two qubits) with $M-j$ qubits in state $|\psi_{\mathrm{in}}\rangle$ and $j$ qubits in the orthogonal state $|\psi_{\mathrm{in}}^{\bot}\rangle \equiv \beta^* |0\rangle - \alpha^*|1\rangle$. More precisely, the \emph{symmetrization operation} as an essential ingredient of the optimal cloning procedure is realized through the action a projection operator $\hat{S}_M$ from an $M$-qubit Hilbert space $\HH^{\otimes M}$ to its completely symmetric subspace $\HH_{S}^{\otimes M}$ given by ($\hat{S}_M: \HH^{\otimes M} \to \HH_{S}^{\otimes M}$)
\begin{eqnarray}
\label{eq:symmetric_state}
|(M-j)\phi,j\phi^{\perp} \rangle_{S} \equiv {1 \over \N} \hat{S}_M |(M-j)\phi,j\phi^{\perp} \rangle  \; ,
\end{eqnarray}
where the prefactor ${1 \over \N}$ accounts for the normalization of the state. As an example, the corresponding projectors for the special cases of $M=1,2$ are, respectively, given by
\begin{subequations}
\label{eq:M_2_symm_projector}
\begin{eqnarray}
\hat{S}_1 & = & |0\rangle \langle 0| + |1\rangle \langle 1| , \\
\hat{S}_2 & = & |00\rangle \langle 00|+|11\rangle \langle 11|+{1 \over 2} (|01\rangle +|10\rangle)(\langle 01|+\langle 10|) \; .
\end{eqnarray}
\end{subequations}
Note that an \emph{a priori} completely symmetric state of $M$ qubits $|\Psi_M\rangle$ remains invariant under the action of the symmetry projector, i.e., $|\Psi_M \rangle_{S} = \hat{S}_M |\Psi_M\rangle= |\Psi_M\rangle$.

The output of the cloning procedure outlined above, i.e., the multiqubit state $|GM_{M}(\psi_{\mathrm{in}})\rangle$, is called the Gisin-Massar state. It describes an entangled state of $M$ identical clones supplemented with $M-1$ anticlones $\psi^{a}_{\mathrm{in}}$ that are introduced to guarantee the optimality of the cloning procedure~\cite{Buzek1996, Gisin1997}. Based on the Bloch sphere representation of qubits, if the cloned states are shown by $|\psi\rangle \equiv \alpha|0\rangle + \beta|1\rangle = \cos{(\theta/2)} |0\rangle + e^{i \phi} \sin{(\theta/2)} |1\rangle$, the state of the corresponding anticlone is given by
\begin{eqnarray}
\label{eq:anti_clone}
|\psi^{a}\rangle \equiv \beta^* |0\rangle + \alpha |1\rangle = e^{-i \phi} \sin{(\theta/2)}|0\rangle + \cos{(\theta/2)} |1\rangle  \; .
\end{eqnarray}

Before embarking on a detailed classical simulation of the Gisin-Massar quantum cloning machine, some remarks on the entanglement characteristics of the output state Eq.~(\ref{eq:Gisin-Massar}) are in order: The Gisin-Massar isometry map $\hat{\V}_{1 \rightarrow M}$ in general represents a non-linear operation in the sense that
\begin{eqnarray}
\label{eq:GM_non_linearity_1}
\hat{\V}_{1 \rightarrow M} (\alpha |\phi_1\rangle + \beta |\phi_2\rangle)\ne \alpha \hat{\V}_{1 \rightarrow M} |\phi_1\rangle + \beta \hat{\V}_{1 \rightarrow M} |\phi_2\rangle \; ,
\end{eqnarray}
or equivalently
\begin{eqnarray}
\label{eq:GM_non_linearity_2}
|GM_{M}(\alpha \phi_1 + \beta \phi_2)\rangle \ne \alpha |GM_{M}(\phi_1)\rangle + \beta |GM_{M}(\phi_2)\rangle \; .
\end{eqnarray}
This implies, in particular, that the cloning of an arbitrary input state $|\psi_{\mathrm{in}}\rangle \equiv \alpha|0\rangle + \beta|1\rangle$ in computational basis $\lbrace |0\rangle,|1\rangle \rbrace$ entails complicated linear combinations of the tensor products of the form
\begin{eqnarray}
\label{eq:GM_decomposed}
\nonumber
|GM_{M}(\alpha 0 + \beta 1)\rangle =\sum_{j=0}^{M-1} \gamma_j \times \hspace{80mm}\\
\nonumber
\Bigl\lbrace |M0\rangle \otimes \Bigl(\alpha \beta^* |\alpha|^2 |(M-1)1\rangle - |\alpha \beta|^2 |(M-j-1)1, j0\rangle_S\\
\nonumber
+ (\alpha \beta^*)^2 |(M-j-1)0, j1\rangle_S - \alpha \beta^* |\beta|^2 |(M-1)0\rangle \Bigr)\\
\nonumber
+ |(M-j)0,j1\rangle_S \otimes \Bigl(-|\alpha|^4 |(M-1)1\rangle + \alpha^* \beta|\alpha|^2 |(M-j-1)1, j0\rangle_S\\
\nonumber
- \alpha \beta^* |\alpha|^2 |(M-j-1)0, j1\rangle_S + |\alpha \beta|^2 |(M-1)0\rangle \Bigr)\\
\nonumber
+ |(M-j)1,j0)\rangle_S \otimes \Bigl( |\alpha \beta|^2 |(M-1)1\rangle - \alpha^* \beta |\beta|^2 |(M-j-1)1, j0\rangle_S\\
\nonumber
+ \alpha \beta^* |\beta|^2 |(M-j-1)0, j1\rangle_S - |\beta|^4 |(M-1)0\rangle \Bigr)\\
\nonumber
+ |M1\rangle \otimes \Bigl(-\alpha^* \beta |\alpha|^2 |(M-1)1\rangle + (\alpha^* \beta)^2 |(M-j-1)1, j0\rangle_S\\
- (|\alpha \beta|^2 |(M-j-1)0, j1\rangle_S + \alpha^* \beta |\beta|^2 |(M-1)0\rangle \Bigr) \Bigr\rbrace\; ,
\end{eqnarray}
where use has been made of an explicit insertion of the input state $|\psi_{\mathrm{in}}\rangle$ in Eq.~(\ref{eq:Gisin-Massar}) and the symmetry property that $(|i\rangle^{\otimes n})_S = |i\rangle^{\otimes n}$ for $i=\lbrace 0,1 \rbrace$.

The key to the classical simulability of the Gisin-Massar state is the important fact that the size of the Hilbert space (and the ensuing computational complexity) associated with such a state does not grow exponentially but rather polynomially with the number of clones $M$. This can be seen by comparing the generic expansion of an $(2M-1)$-qubit state in the computational basis of the form
\begin{eqnarray}
\label{eq:GM_comp_basis}
|GM_M(\psi)\rangle = \sum_{i_1=0}^1 \sum_{i_2=0}^1 \dots \sum_{i_{2M-1}=0}^1 c^{\psi}_{i_1, i_2, \cdots ,i_{2M-1}}
|i_1, i_2, \dots ,i_{2M-1}\rangle   \; ,
\end{eqnarray}
involving a coefficient tensor $c^{\psi}$ of rank $(2M-1)$ to that of the expansion Eq.~(\ref{eq:Gisin-Massar}) with a single index $j$ controlling the number of terms in the corresponding expansion of the multiqubit output state.
More precisely, the symmetrization requirement of the Gisin-Massar state within either clone or anticlone subspace establishes the algebraic connection between $\gamma_j$'s and $c^{\psi}$'s. It indeed makes most of the coefficients $c^{\psi}$ vanish or to be identical. Distinct values of $\gamma_j$ correspond only to \emph{a priori} symmetrized products of clone kets and the anticlone ones (those contributions without the symmetrization $S$ index in Eq.~(\ref{eq:GM_decomposed})) and are equal in number to the number of clones $M$ to be produced. It has been demonstrated numerically by the present authors in Ref.~\cite{Saberi2012} that such a feature is responsible for the identification of the Gisin-Massar state as the one belonging to the important class of \emph{slightly entangled} multiqubit states~\cite{Vidal2003} with the possibility to be simulated efficiently on a classical computer.

Realizing the quantum cloning map $\hat{\V}_{1 \rightarrow M}$, though, through a single application of a global isometry operation that entangles all input qubits simultaneously is in general a task of formidable difficulty. A more viable scenario in terms of the physical implementation of the Gisin-Massar quantum cloning map was put forward by Delgado $\emph{et al}$~\cite{Delgado2007,Lamata2007} and is based on a \emph{sequential} implementation of the cloning procedure. In the sequential paradigm of quantum cloning, an ancillary system is introduced to interact locally and only once with each qubit in a row, mediating thereby through engineerable ancilla-qubit interactions the desired form of entanglement among the qubits, and is set to eventually decouple from the qubit chain in the last step.


On the other hand, the output from such a sequentially implemented quantum cloner can be characterized in terms of the hierarchy of the so-called matrix-product states (MPS)~\cite{Perez2007,Verstraete2008} as the one-dimensional version of the well-studied class of tensor networks~\cite{Saberi2008, Weichselbaum2009, Weichselbaum2012}. An MPS representation of the $(2M-1)$-qubit output of the Gisin-Massar state is given by
\begin{eqnarray}
\label{eq:MPS_GM}
|{GM}_M(\psi)\rangle = \sum_{i_{2M-1} \dots i_1=0}^1 \langle\varphi_F| A_{[2M-1]}^{i_{2M-1}} \cdots A_{[1]}^{i_1} |\varphi_I\rangle |i_{2M-1},\cdots,i_1\rangle
\; ,
\end{eqnarray}
where $(D_k\times D_{k+1})$-dimensional matrix $A_{[k]}^{i_k}$ represents the physical interaction between ancilla and $k$th qubit with local space $|i_k\rangle$, and with $|\varphi_I\rangle$ and $|\varphi_F\rangle$ denoting the initial and final ancilla state, respectively. The \emph{bond dimension} of an MPS is defined then as $D \equiv \mathrm{max}_k D_k$.

According to the Delgado \emph{et al}., the minimal bond dimension of the MPS representation (\ref{eq:MPS_GM}) coincides with the required ancilla dimension $D$ (e.g., the number of atomic levels) for sequential implementation of the desired Gisin-Massar quantum cloner~\cite{Delgado2007}. Such an analysis also revealed a \emph{linear} scaling of the required minimal ancilla dimension $D$ with the number of clones $M$ (more precisely $D=2M$ for the case of universal symmetric cloning of Gisin-Massar as described in the subsequent section). Further numerical analysis by the present authors yet clarified an almost constant scaling of the ancilla dimension with the number of clones up to 15 qubits~\cite{Saberi2012}. The latter promises a classically feasible simulation of the cloning scheme within the framework of the tensor network formalism. It is the purpose of the present paper to collect the technical details of such a computational protocol.

In this work, we focus the cloning of a completely symmetric input state of the form $|\psi_{\mathrm{in}}\rangle = \frac{1}{\sqrt{2}}( |0\rangle + |1\rangle )$ as a state located on the equator of the Bloch sphere and categorized as an \quot{equatorial} qubit~\cite{Fan2001} defined by a pure superposition of the form $\alpha |0\rangle + \beta |1\rangle$ with real-valued coefficients $\alpha$ and $\beta$ and the normalization condition $\alpha^2 + \beta^2 = 1$. The procedure associated with the quantum cloning of such a restricted set of input states is called \quot{phase-covariant} quantum cloning~\cite{Bruss2000}. Note that the Bloch vector for such states is restricted to the intersection of the $x$-$y$ plane with the Bloch sphere. Alternatively, one could write the equatorial qubit as $|\psi_{\mathrm{eq.}}\rangle = {1 \over \sqrt{2}} (|0\rangle + e^{i \phi} |1\rangle)$ entailing a single unknown independent real parameter $\phi$ that can be identified as the angle between the Bloch vector and the $x$ axis.)

 In the following section, we provide the technical details for obtaining an MPS representation of a Gisin-Massar state to set the scene for the description of its practical implementation on classical computers throughout Sec.~\ref{sec:simulation_protocol}.

\section{Matrix-product state representation of the Gisin-Massar state}
\label{sec:MPS_GM}

In this section we elaborate on a basic technique for obtaining an MPS representation of the original Gisin-Massar state (\ref{eq:Gisin-Massar}) by performing successive singular-value decomposition (SVD) on all possible bipartite decompositions of the coefficient tensor $c^{\psi}_{i_{2M-1}, i_{2M-2}, \cdots, i_1}$ in Eq.~(\ref{eq:GM_comp_basis}). To this end, we start by a bipartite decomposition of the coefficient indices as $i_1|i_2 \dots i_{2M-1}$ reshaping thereby the coefficient tensor $c^{\psi}$ of rank $2M-1$ into a new \emph{matrix} $C^{(1)}$ with respective indices $c_{i_1, \underline{i_2 i_3 \cdots i_{2M-1}}}$ in which $\underline{i_2 i_3 \cdots i_{2M-1}}$ is being treated as a single and \quot{coarse-grained} combined super-index. Performing SVD on ${C}^{(1)}$ yields
\begin{eqnarray}
\label{eq:SVD_C1}
{C}^{(1)} = V^{(1)} S^{(1)} {W^{(1)}}^{\dagger} \; ,
\end{eqnarray}
where $V^{(1)} (W^{(1)})$ denotes the left (right) unitary in the decomposition and $S^{(1)}$ is the diagonal matrix containing the singular values of the original matrix $C^{(1)}$. Component-wise, we may then write the latter as
\begin{eqnarray}
\label{eq:SVD_C1_component}
c_{i_1, \underline{i_2\cdots i_{2M-1}}}= \sum_{k} v_{i_1 k}^{(1)} {(S^{(1)} {{V}^{(1)}}^{\dagger})}_{k, \underline{i_2\dots i_{2M-1}}} \; .
\end{eqnarray}
Define now the matrix
\begin{eqnarray}
\label{eq:A1}
A_{[1]} \equiv V^{(1)} S^{(1)}  \; ,
\end{eqnarray}
with its $i_1$th row representing the sub-matrices $A_{[1]}^{i_1}$ associated with the local space of the first qubit. We proceed by performing another SVD now on the remaining part ${W^{(1)}}^{\dagger}$ considering this time the partitioning $\underline{i_1 i_2}|\underline{i_3 \cdots i_{2M-1}}$ as follows
\begin{eqnarray}
\label{eq:SVD_C2}
C^{(2)} \equiv {W^{(1)}}^{\dagger} = V^{(2)} S^{(2)} {W^{(2)}}^{\dagger} \; ,
\end{eqnarray}
or component-wise
\begin{eqnarray}
\label{eq:SVD_C2_component}
({W^{(1)}}^{\dagger})_{i_1, \underline{i_2 \cdots i_{2M-1}}} \equiv c_{\underline{i_1 i_2},\underline{i_3 \cdots i_{2M-1}}}= \sum_{k} v_{\underline{i_1 i_2}, k}^{(2)} {(S^{(2)} {{V}^{(2)}}^{\dagger})}_{k, \underline{i_3\dots i_{2M-1}}} \; ,
\end{eqnarray}
to identify again
\begin{eqnarray}
\label{eq:A2}
A_{[2]} \equiv V^{(2)} S^{(2)}  \; ,
\end{eqnarray}
with its $\underline{i_1 i_2}$th rows representing the sub-matrices $A_{[2]}^{i_2}$ associated with the local space of the second qubit. Iterating such a procedure till the end of the qubit chain gives rise to a matrix-product decomposition of the original coefficient matrix of the form
\begin{eqnarray}
\label{eq:MPS_coeff}
c^{\psi}_{i_{1}, i_{2}, \cdots, i_{2M-1}} = A_{[1]}^{i_1} A_{[2]}^{i_2} \cdots A_{[2M-1]}^{i_{2M-1}} \; .
\end{eqnarray}

Although the cloning of an arbitrary input state can not be decomposed into the separate cloning of the computational kets $|0\rangle$ and $|1\rangle$ owing to the nonlinearity of the Gisin-Massar map detailed in Section~\ref{eq:Gisin-Massar}, but the paradigm of sequential quantum cloning is capable of realizing such a task by explicit construction of the respective MPS representation of the Gisin-Massar outputs for $|0\rangle$ and $|1\rangle$ given by
\begin{subequations}
\label{eq:MPS_0_1}
\begin{eqnarray}
|{GM}_M(0)\rangle = \sum_{i_{1} \cdots i_{2M-1}} \langle\varphi_I| A_{0[1]}^{i_1} \cdots A_{0[2M-1]}^{i_{2M-1}} |\varphi_F\rangle |i_1,\cdots,i_{2M-1}\rangle , \\
|{GM}_M(1)\rangle = \sum_{i_1 \cdots i_{2M-1}} \langle\varphi_I| A_{1[1]}^{i_1} \cdots A_{1[2M-1]}^{i_{2M-1}} |\varphi_F\rangle |i_1,\cdots, i_{2M-1}\rangle \; ,
\end{eqnarray}
\end{subequations}
and doubling thereby the ancilla dimension to account for the generation of the respective isometries $A_{0[i_k]}$'s or $A_{1[i_k]}$'s~\cite{Delgado2007}.

    Despite the straightforward way from the computational basis representation of the Gisin-Massar state to the one in MPS form outlined above, the preparation step for transforming the original form of the Gisin-Massar [Eq.~(\ref{eq:Gisin-Massar})] into a classically programmable way within a complete orthonormal basis is not a trivial task to accomplish. Much of the complication in that respect arises from the fact that it is in general a tedious task to perform all the required symmetrization operations of Gisin-Massar for arbitrary input states. However, the situation improves upon expressing the states in the computational basis in terms of classical bit (cbit) variants ${0,1}$ on a classical computer. In the subsequent section, we provide an efficient protocol for preparation of the state of Gisin-Massar from in computational basis without struggling with the symmetrization requirements.

\section{Classical implementation of the Gisin-Massar quantum cloner}
\label{sec:simulation_protocol}

The first step in simulation of the Gisin-Massar state of the form Eq.~({\ref{eq:Gisin-Massar}}) is the preparation of the multiqubit state on cbits of a classical computer. This essentially corresponds to feed the proper input state in computational basis for the sake of obtaining the MPS representation required for the sequential paradigm of quantum cloning. Besides the computational cost incurred upon performing the symmetrization operations in the original representation of the Gisin-Massar state [Eq.~(\ref{eq:Gisin-Massar})], other practical complications and ambiguities arise as a result of the action of the symmetrization operators: The coefficients $\gamma_j$ there exhibit high degeneracies in that so many distinct computational basis kets $|i_1,i_2,\cdots,i_{2M-1}\rangle$ correspond to the same coefficient $\gamma_j \equiv c_{i_1,i_2,\cdots,i_{2M-1}}^{(\psi)}$. In fact, the total number of distinct values of coefficients equals only $M$ out of (instead of) the maximal possible value $2^{2M-1})$ for a generic $|GM_M(\psi)\rangle$. In other words,
such a degeneracy causes indistinguishability between the Gisin-Massar eigenstates making an \emph{in situ} discrimination of the kets hard to realize.
The degeneracy, however, can be lifted upon noticing an important and fortunate parity feature about the output clones of $|0\rangle$ and $|1\rangle$
\begin{subequations}
\label{eq:GM_0_1}
\begin{eqnarray}
|GM_M(0)\rangle = \sum_{j=0}^{M-1} \gamma_{j}|(M-j)0,j1\rangle \otimes |(M-j-1)1,j0\rangle_{S} , \\
|GM_M(1)\rangle = \sum_{j=0}^{M-1} \gamma_{j}|(M-j)1,j0\rangle \otimes |(M-j-1)0,j1\rangle_{S} \; ,
\end{eqnarray}
\end{subequations}
as follows: The total number of qubits in state $|1\rangle$ for the clone of state $|0\rangle$ reads $M-1$ whereas it turns out to be $M$ for the clone of state $|1\rangle$. Defining a parity operator $\hat {\cal{P}}$ whose action on a register of qubits with an \emph{even} (\emph{odd}) number of qubits in state $|1\rangle$ yields an eigenvalue $+1 (-1)$, then the desired distinction between those contributions pertaining to the clone of $|0\rangle$ or $|1\rangle$ can be realized easily depending on wether the number of clones $M$ is odd or even.

The outlined procedure may be found analogous to the common parity bit identification technique which has been widely used in many areas of classical computation and information such as error-correction, classical and quantum cryptography~\cite{Bennett1996}, the backup and recovery specially in redundant array of independent disks (RAIDs)~\cite{Patterson1987}, and also fault-tolerant optical quantum computing schemes~\cite{Hayes2010}.

All in all, the proposed algorithm for the state preparation on a classical computer can be stated as follows:

1. One generates first all possible permutations of digits $\{0, 1\}$ for a bit sequence of length $2M-1$. The sorted sequences can be saved then on a text file named say $\mathrm{\mathbf{FullBitString}}$.

2. One produces afterward the Gisin-Massar kets sequences associated with the clones of $|0\rangle$ and $|1\rangle$ using \Eq{eq:GM_0_1}, sorts them and saves on a text file named say $\mathrm{\mathbf{GMBitString}}$.

3. A proper string search algorithm is utilized to extract those Gisin-Massar contributions that are contained within $\mathrm{\mathbf{FullBitString}}$ for which the corresponding $\gamma_{j}$ will be calculated from Eq.~{(\ref{eq:gamma_j})} with the outcome written on another text file named say $\mathrm{\mathbf{GMMatrix}}$. Next the right $\gamma_j$ shall be assigned to the right Gisin-Massar ket. The above-mentioned ambiguity in making the right distinction arises right here. However, as described above, the issue can be circumvented upon exploiting the parity feature of Gisin-Massar kets by simply counting the 1's in each sequence of digits to figure out if they pertain to the clone of state $|0\rangle$ or $|1\rangle$.

Once formed the coefficient matrix $\mathrm{\mathbf{GMMatrix}}$, one can routinely proceed with obtaining the corresponding MPS representation following the recipe of the preceding section.

The proposed protocols have successfully been utilized by the present authors for the sake of addressing real-life experimental challenges associated with the physical realization of a sequential \quot{factory} of quantum cloning~\cite{Saberi2012}.

\section{Conclusions and outlook}
\label{sec:conclusions}

In conclusion, we have provided practical recipes for an efficient simulation of the optimal quantum cloning procedure of Gisin-Massar. The protocol instructs a classical computer how to prepare the Gisin-Massar entangled state on cbits and transform the outcome into an MPS representation suitable for the purpose of a sequential and scalable implementation of the machine in a form amenable to the potentially scalable physical setups either in optical systems
~\cite{Zhao2005,Bartuskova2007,Nagali2009} or NMR setups~\cite{Cummins2002,Du2005,Chen2007}.

No-cloning theorem plays an important role in the context of quantum cryptography where it is responsible for the security of quantum key distribution (QKD) protocols upon neutralizing the quantum cloning attacks by an eavesdropper. As such, quantum cloning machines can be designed to analyze the security of QKD protocols such as the well-accepted Bennett-Brassard 1984 (BB84) quantum cryptography protocol~\cite{Bennett1984} employing four equatorial quantum states to transmit information through a public quantum channel.

Moreover, we expect that the parity techniques introduced in this work could well be invoked in the context of entanglement-by-measurement and economical projective parity measurements where a \emph{qubit parity} determines whether an even or odd number of qubits is in a particular eigenstate~\cite{Pfaff2012}. An ideal ancilla-assisted \emph{parity measurement} followed by a high-fidelity readout of the ancillary system projecting the state of the qubits register into the pertinent subspace with either even or odd parity. A heralded qubit parity measurement on two nuclear spins in diamond by using the electron spin of a nitrogen-vacancy defect center as a readout ancilla has recently been reported in Reference~\cite{Pfaff2012}.
\\

The authors gratefully acknowledge stimulating discussions with Enrique Solano and Lucas Lamata.
H.S. acknowledges support from the vice-president office for research and technology affairs of Shahid Beheshti University, and
the German Research Foundation (DFG) under SFB 689. H.S. is also grateful to Universidad del Pa\'{\i}s Vasco for support and hospitality.





\bibliographystyle{elsarticle-num}
\bibliography{Gisin-Massar}







\end{document}